\documentclass{article}

\PassOptionsToPackage{numbers, compress}{natbib}

\usepackage[final]{neurips_2023_ml4ps}
\usepackage{graphicx}    
\usepackage{amsmath}    
\usepackage{tabulary}
\usepackage{enumitem}
\usepackage{footnote}
\usepackage{hyperref}
\usepackage{natbib}
\usepackage{caption}
\usepackage{subcaption}
\usepackage{fontawesome}
\usepackage{graphicx}
\usepackage{bm}
\usepackage[dvipsnames]{xcolor}

\usepackage[utf8]{inputenc} 
\usepackage[T1]{fontenc}    
\usepackage{hyperref}       
\usepackage{url}            
\usepackage{booktabs}       
\usepackage{amsfonts}       
\usepackage{nicefrac}       
\usepackage{microtype}      
\usepackage{xcolor}         

\title{
Extracting an Informative Latent Representation of High-Dimensional Galaxy Spectra
}

\author{%
    \href{mailto:iwasaki.daiki.p6@s.mail.nagoya-u.ac.jp}{Daiki Iwasaki}$^{1}$, 
    \href{mailto:cooraysuchetha@gmail.com}{Suchetha Cooray}$^{2}$, 
    and \href{mailto:tsutomu.takeuchi.ttt@gmail.com}{Tsutomu T. Takeuchi}$^{1,3}$ \\
  $^{1}$ Nagoya University \\
  $^{2}$ National Astronomical Observatory of Japan \\
  $^{3}$ Institute of Statistical Mathematics \\
}

\begin{document}

\maketitle

\begin{abstract}
To understand the fundamental parameters of galaxy evolution, we investigated the minimum set of parameters that explain the observed galaxy spectra in the local Universe. We identified four latent variables that efficiently represent the diversity of high-dimensional galaxy spectral energy distributions (SEDs) observed by the Sloan Digital Sky Survey. Additionally, we constructed meaningful latent representation using conditional variational autoencoders trained with different permutations of galaxy physical properties, which helped us quantify the information that these traditionally used properties have on the reconstruction of galaxy spectra. 
The four parameters suggest a view that complex SED population models with a very large number of parameters will be difficult to constrain even with spectroscopic galaxy data. 
Through an Explainable AI (XAI) method, we found that the region below 5000\textup{\AA} and prominent emission lines ([O II], [O III], and H$\alpha$) are particularly informative for predicting the latent variables. Our findings suggest that these latent variables provide a more efficient and fundamental representation of galaxy spectra than conventionally considered galaxy physical properties.

\end{abstract}


\section{Introduction} \label{Introduction}
A galaxy's spectral energy distribution (SED), which reflects the multi-wavelength flux observed of a galaxy, is often the only path to study the astrophysical processes within them. 
Instruments like the current Dark Energy Spectroscopic Instrument (DESI), upcoming Prime-Focus Spectrograph (PFS) instrument, and integral field units (IFUs) like the Multi Unit Spectroscopic Explorer (MUSE), obtain high-dimensional data for sometimes millions of galaxies, posing computational complexity challenges and issues inherent to high-dimensional analysis, such as overfitting and multicollinearity. Traditionally, we have employed various methods that make high-dimensional data manageable, such as photometry and color-magnitude diagrams \citep[e.g.,][]{bell_nearly_2004} and the BPT diagram \citep{baldwin_classification_1981}, to characterize galaxy spectra based on their representative low dimensional features. However, these low-dimensional representations do not contain the complete spectral information. Hence, we use neural networks to extract full information from high-dimensional galaxy spectroscopic data. In doing so, this work addresses three main scientific questions: 
1. How many parameters are required for adequate representation of galaxy SEDs? 
2. What fundamental physical properties explain the observed SEDs? 
3. Which spectral ranges are the most informative for representing observed SEDs?

This work uses the Variational Autoencoder \citep[VAE;][]{kingma_auto-encoding_2013}, a type of neural network to compress galaxy spectra into meaningful, lower-dimensional latent parameters without relying on explicit labels or prior assumptions. To identify the most informative physical properties in reconstructing galaxy spectra, we employ the Conditional VAE \citep[CVAE;][]{kingma_semi-supervised_2014}, incorporating conditional properties such as stellar mass (\( M_* \)), star formation rate (SFR), specific SFR (sSFR = SFR/\( M_* \)), and metallicity ($Z$). We also utilize the XAI method named SHAP \citep[SHapley Additive exPlanations;][]{lundberg_unified_2017} to assign importance values to specific spectral features in galaxy SEDs. This enables us to identify key spectral ranges that influence the VAE's latent variables, providing insight into the fundamental characteristics of galaxies.

\section{Data and Methodology} \label{Data and Methodology}
The primary dataset for this analysis is the galaxy spectroscopic data from the Sloan Digital Sky Survey 
 \citep[SDSS;][]{ahumada_16th_2020}. Physical properties were derived from the GALEX-SDSS-WISE Legacy Catalog \citep[GSWLC;][]{salim_galex-sdss-wise_2016}, which used UV+optical+mid-IR SED fitting. Our final dataset includes SDSS spectra cross-matched with the GSWLC. We limit to galaxies with a redshift < 0.1 to ensure consistency in the spectral range across our dataset. For preprocessing spectra, we used the Python module \texttt{spectres} to shift each galaxy spectrum to its rest frame and to resample it to 4000 logarithmically spaced wavelength pixels within the range of 3400\AA–8400\AA. A similar approach was employed in previous studies \citep[e.g.,][]{portillo_dimensionality_2020,pat_reconstructing_2022}. We also used an iterative principal component analysis (PCA) method to handle bad data points, following a similar approach used by \citep{yip_distributions_2004}. The dataset utilized for this study comprises a total of about 320,000 galaxy spectra.



We employ a VAE and its variant CVAE for this analysis. VAE is a two-component architecture, an encoder and a decoder, where the encoder maps high-dimensional data into a usually lower-dimensional latent space, and the decoder reconstructs the original input from lower-dimensional latent variables. The VAE loss function is the sum of two components: the reconstruction error, which is the distance metric between the input data and the reconstructed output and Kullback-Leibler divergence \citep{kullback_information_1951}, which encourages the latent variables to usually follow a standard Gaussian distribution, $N(0,1)$. The VAE achieves a continuous lower-dimensional latent representation of the complex, high-dimensional data by optimizing this composite loss function(refer to the Appendix \ref{VAE}). CVAE integrates conditional data during the encoding and decoding processes, enabling more controlled output generation while retaining the same loss function as the VAE. By utilizing physical properties as conditional data, we obtain latent representations that remain unaffected by them. 

The deep neural network model, often seen as a "black box," poses challenges in understanding the underlying reasons for its predictions. To address this problem, we utilized SHAP values, which quantify the influence of each feature on the model's predictions, thereby elucidating the impact of different inputs on the outcome (refer to the Appendix \ref{The SHAP values analysis}). All scripts for downloading SDSS spectra, preprocessing data, training, and generating the figures are accessible \faGithub\href{https://github.com/iwasakida}{here}.

\section{Results}
\begin{figure*}
    \centering
        \begin{minipage}{0.48\textwidth}
            \centering
            \includegraphics[width=\textwidth]{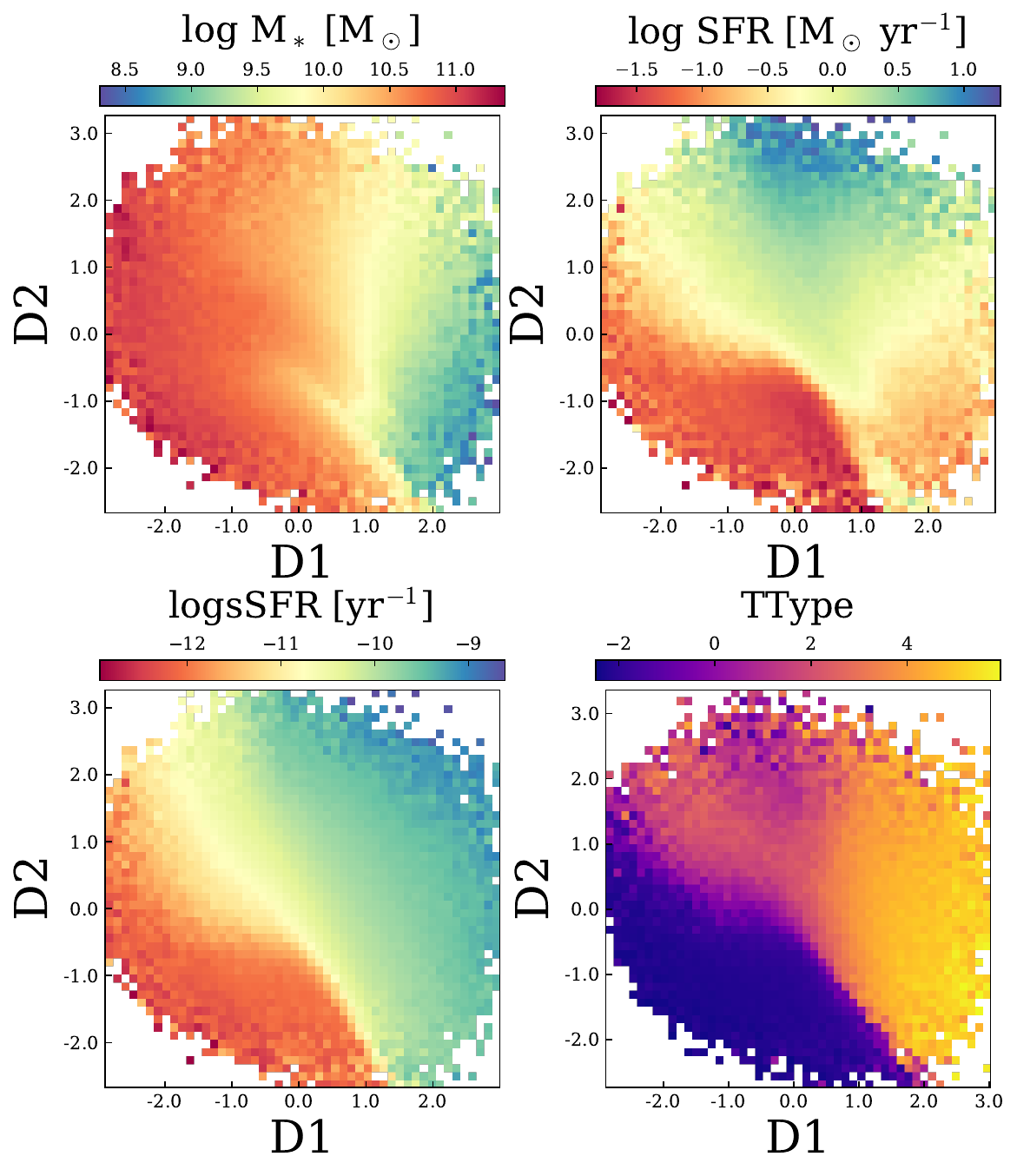}
            \caption{
            Distribution of average stellar masses, SFR, sSFR, and TType in the latent space. The figures from the top right to the bottom left display the distributions of $M_{*}$, SFR, sSFR, and TType in the latent space. Median values for each property are calculated within their respective bins. The morphological properties are derived from \citealp{dominguez_sanchez_improving_2018}.
            }
            \label{figure physical property in latent space}
        \end{minipage}\hfill
        \begin{minipage}{0.5\textwidth}
            \centering
            \includegraphics[width=\textwidth]{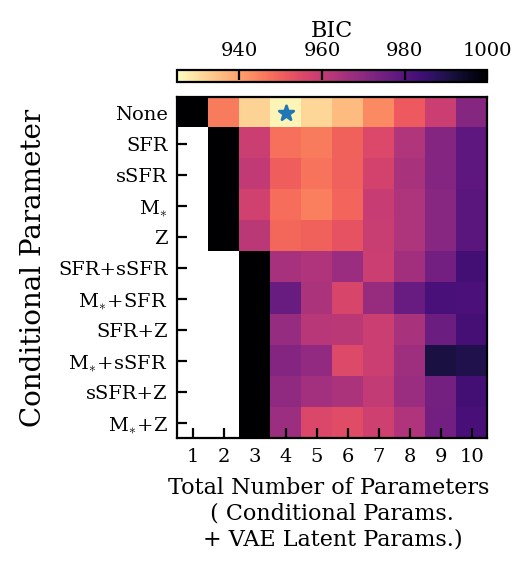}
            \caption{
            BIC values for combinations of latent variables and conditional data. BIC values are presented as a function of model complexity, with total parameters on the x-axis and conditional data on the y-axis. The color gradient depicts BIC values: darker shades represent higher values, and lighter shades indicate lower ones. A model with a lower BIC is typically a better fit. The blue star marks the model that achieves the optimal balance between fit and complexity. 
            }
            \label{figure BIC}
        \end{minipage}
\end{figure*}

The primary results of our study indicate that four latent variables can effectively represent the 320k, 4000-dimensional galaxy spectra, showing strong correlations between the observed diversity of galaxy physical properties. These latent parameters also show correlations with traditional galaxy physical properties such as stellar masses, star formation rates, specific star formation rates, and metallicity, as shown in Figure \ref{figure physical property in latent space}.


We use the Bayesian information criterion \citep[BIC;][]{schwarz_estimating_1978} to identify the optimal number of parameters for spectral reconstruction. The BIC is a criterion used for model selection among models with varying permutations of conditional parameters. BIC is defined as $\text{BIC} = k\log(n) - 2\log(L)$ where $k$ is the total number of parameters and $L$ is the likelihood. BIC, derived from Bayesian principles, aims to identify the 'true' model among the candidates. This approach is distinct from other criteria based on information entropy \cite{akaike_new_1974}. Figure \ref{figure BIC} presents the BIC values as a function of the total number of parameters (latent variables + conditional data). We successfully quantified the extent to which physical properties influence the reconstruction of galaxy spectra and the capture of an informative latent representation. Our results indicate that the VAE with four parameters is the best model. The fact that the VAE outperforms the CVAE suggests that the latent variables obtained provide a more efficient representation for characterizing galaxy spectra distribution compared to traditional galaxy physical properties. We also found that increasing the number of conditional parameters does not improve the results significantly. Moreover, when excluding the VAE, the CVAE model conditioned with $M_{*}$ and the model conditioned with SFR performs well under our metric. The former yielded a result of 945.67, while the latter scored 945.14, indicating a difference of just approximately 0.5. The standard deviation of the BIC values for models with five parameters is 8.83, so this 0.5 difference is not very significant. This result is equivalent to what would be obtained by determining the Evidence Lower Bound with an added KL divergence loss term.
\section{Discussion} \label{discussion}
\subsection{How many parameters are required for adequate representation of galaxy SEDs?} 
The latent variables D${_1}$ to D${_4}$ are PCA-transformed values of the VAE latent features. PCA is a lossless transformation that preserves information because it does not alter the number of parameters \citep[][]{pearson_liii_1901}. In the original VAE latent space, axes are arbitrary and may not correspond to meaningful directions. By applying PCA tranformation to the latent variables, the latent axes are the most informative. A similar approach was done in previous studies \citep[]{portillo_dimensionality_2020,pat_reconstructing_2022}. D${_1}$ to D${_4}$ capture 34\%, 32\%, 16\%, and 18\% of the total variance, respectively.

\begin{figure*}
    \begin{minipage}{0.45\textwidth}
        \centering
        \includegraphics[width=\textwidth]{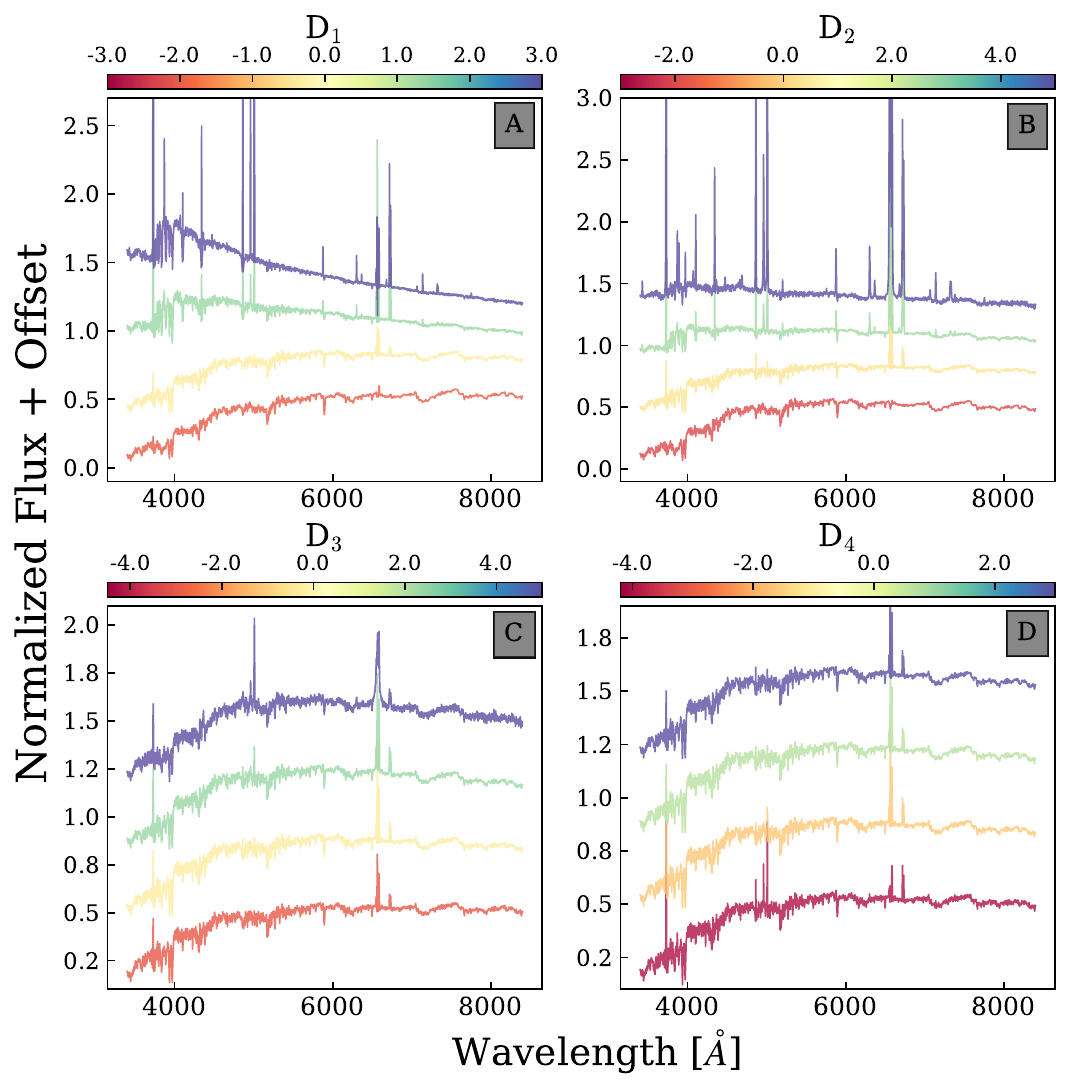}
        \caption{
        Effect of single latent variable change on reconstructed spectra. With all other parameters held at zero, the figure demonstrates the influence of varying just one latent variable on the generated spectra. The four panels, from A to D, correspond to different latent variables: D${_1}$ through D${_4}$. 
        }
        \label{figure spectra change}
    \end{minipage}\hfill
    \begin{minipage}{0.5\textwidth}
        \centering
        \includegraphics[width=0.9\textwidth]{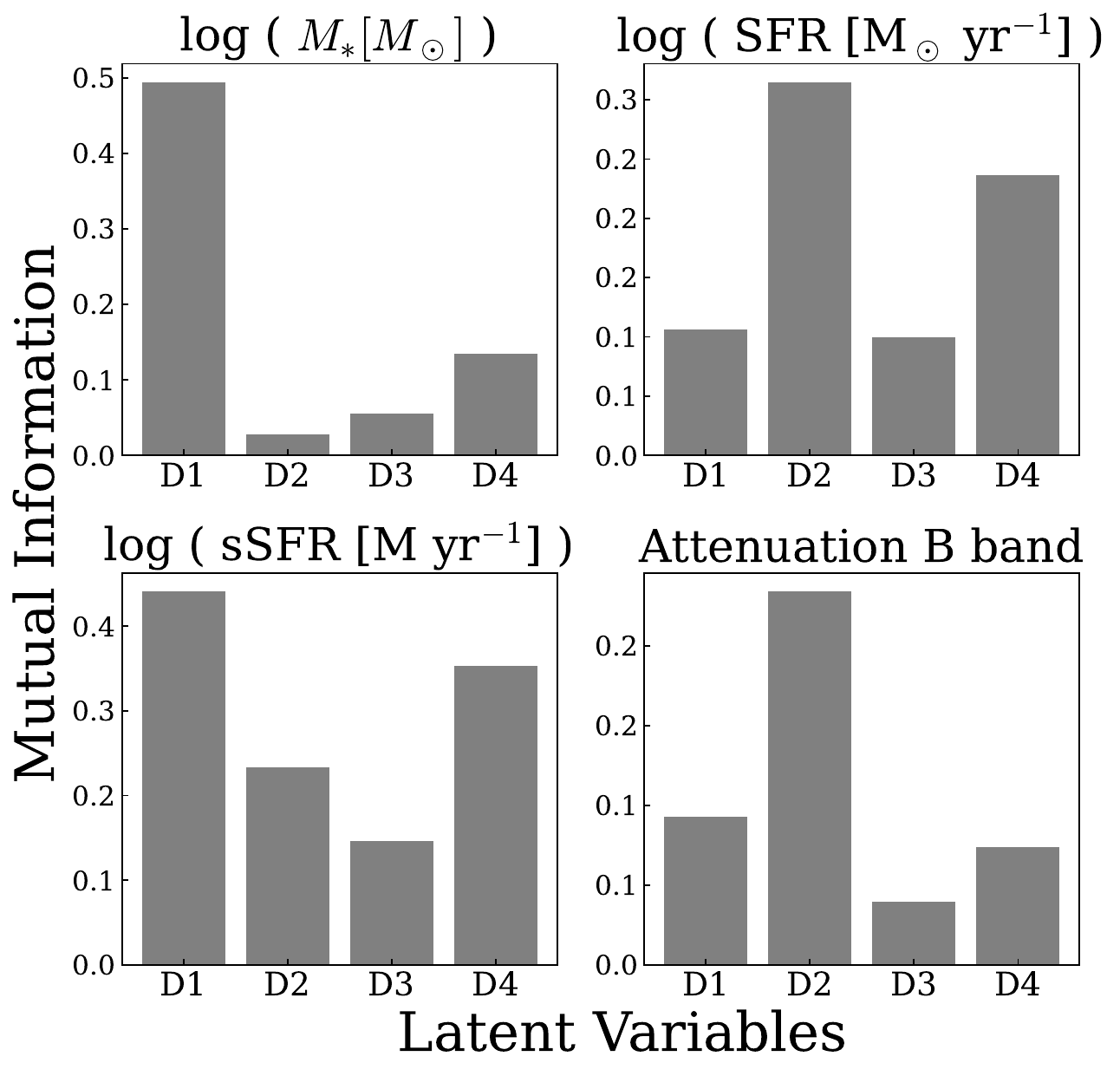}
        \caption{
             Mutual information between parameters and latent variables (D${1}$ to D${4}$). These eight bar plots depict the MI between parameters and latent variables (D${1}$ to D${4}$). Parameters include $M_{*}$, SFR, sSFR, Metallicity, exponential $\tau$, Velocity Dispersion, Attenuation in B band, and V band, ordered top left to bottom right. The y-axis signifies MI value, and the x-axis presents the four latent variables.
        }
        \label{figure mutual information}
    \end{minipage}\hfill
\end{figure*}

Figure \ref{figure spectra change}, shows the effect of changing the latent parameters on the reconstructed spectra. We used Mutual Information (MI) values to understand the link between latent variables and physical properties (Figure \ref{figure mutual information}). As seen in Figure \ref{figure spectra change} Panel A, D$_{1}$ impacts the 4000\textup{\AA} break, signifying evolving stellar populations along D$_1$. Higher D$_{1}$ aligns with star-forming galaxies, while lower values indicate older, quiescent galaxies. D$_1$ also has the highest correlation with $M_{*}$, then sSFR, and lastly, metallicity. This reflects a trend from lower-mass to massive galaxies along D$_1$ as seen in Figure \ref{figure physical property in latent space}. Modifying D$_{2}$ (Figure \ref{figure spectra change}, Panel B) alters the spectral intensity, mainly below 5000\textup{\AA}, while increasing emission line strength to around 6000\textup{\AA}. Figure \ref{figure mutual information} shows D$_2$ exhibits a strong correlation with SFR. D$_1$ and D$_2$ account for roughly 66\% of the total variance. The low MI values between D$_1$ and SFR, and D$2$ and $M{*}$, combined with improved reconstructions using SFR and M as data, emphasize the importance of $M{*}$ and SFR in describing galaxy spectra. These patterns remained consistent across different number of latent variables settings, underlining the key roles of D$_1$ with $M{*}$ and D$_2$ with SFR in galaxy spectrum characterization.

Figure \ref{figure spectra change} (Figure \ref{figure spectra change}, Panel C and D) displays how D$_{3}$ and D$_{4}$ influence emission line intensities like [O II], [O III], and H$\alpha$ without altering the stellar continuum. D$_3$ correlates with ionized gas attributes and Active Galactic Nuclei (AGNs) presence, especially the [O III] emission line. Galaxies with AGNs or ionization from evolved stellar populations exhibit pronounced [O III] lines and higher D4000 values, indicating a larger fraction of older stars. While D$_3$ shows variable [O III] intensity, H$\alpha$ remains stable. In D$_4$, oxygen lines and H$\alpha$ increase simultaneously. H$\alpha$ from ionized hydrogen gas, indicates active star formation, with its strength correlating with the [O II] emission due to ionization by young stars. However, AGNs can alter this relationship. 

To more carefully interpret these latent representations, we use SHAP values to evaluate the significance of individual input wavelengths for predicting latent variables. A positive SHAP value indicates that increasing the feature's value increases the prediction, while a negative value decreases. In Figure \ref{figure SHAP value} Panel A, SHAP value of D${_3}$ peaks at [O II] emission lines and declines at [O III] and H$\alpha$ emission lines. Similarly, Panel B shows SHAP value of D${_4}$ rises at both [O II] and [O III] emission lines and falls at H$\alpha$ emission lines. Therefore, we conclude that D${_3}$ represents AGN strength, signifying the ratio of [O II] strength compared to [O III] and H$\alpha$, while D${_4}$ indicates the ratio of H$\alpha$ strength relative to [O III] and [O II] emission lines. 
This study's results are based on a single dataset and single model. There is a need for further investigation using additional datasets to verify and extend these findings.

\subsection{What fundamental physical properties explain the observed SEDs?}
\begin{figure*}
    \begin{minipage}{0.49\textwidth}
        \centering
        \includegraphics[width=\textwidth]{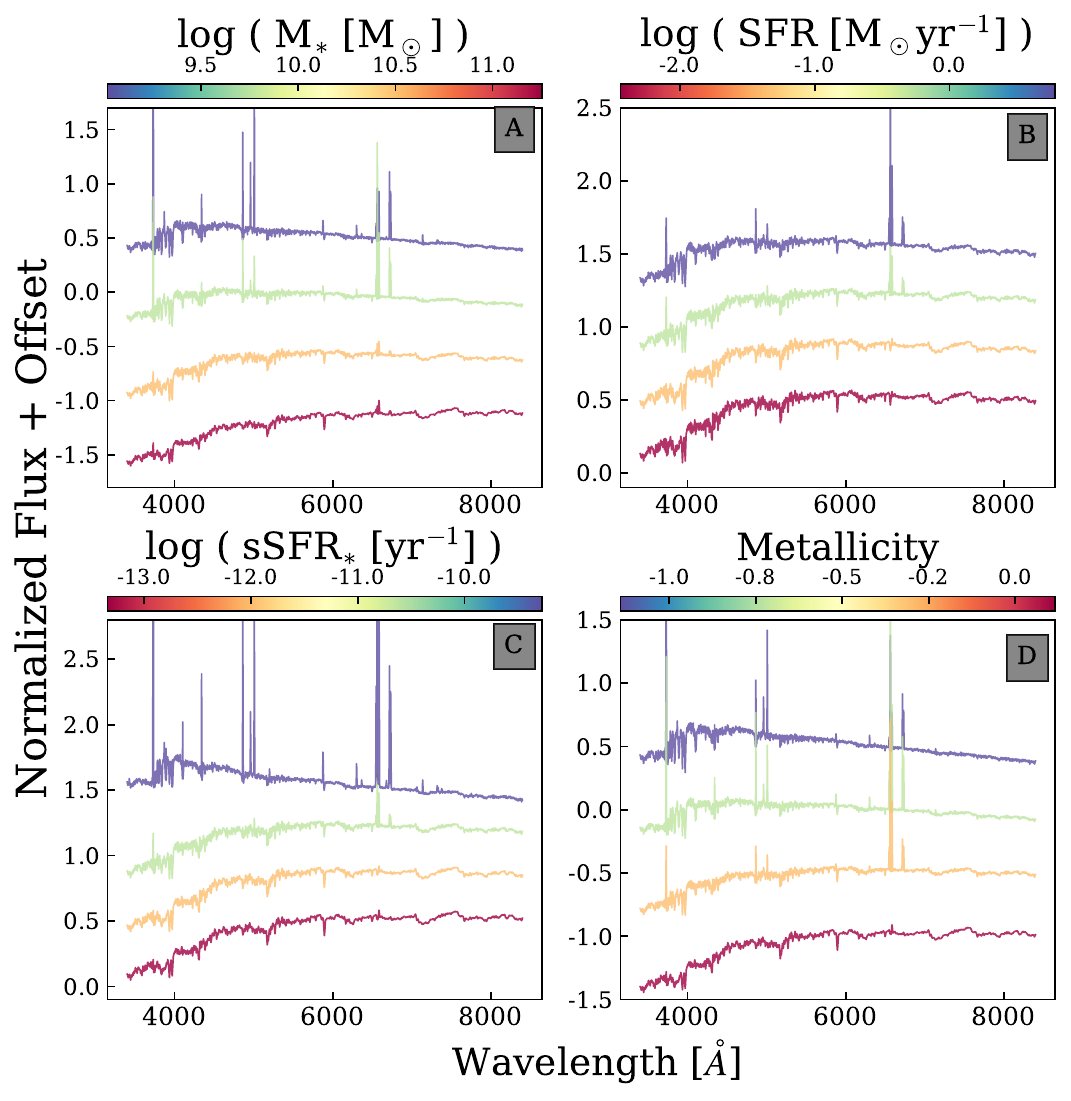}
        \caption{
        This figure consists of four panels (labeled A to D), each representing a different conditional data parameter: SFR, M, sSFR, and Metallicity. The four spectra within each panel are generated by a CVAE model, with one latent parameter held at zero while varying the values of the corresponding conditional parameter. The color of each spectrum represents the value of the respective conditional parameter.
        }
        \label{figure spectra change CVAE}
    \end{minipage}\hfill
    \begin{minipage}{0.49\textwidth}
        \centering
        \includegraphics[width=\textwidth]{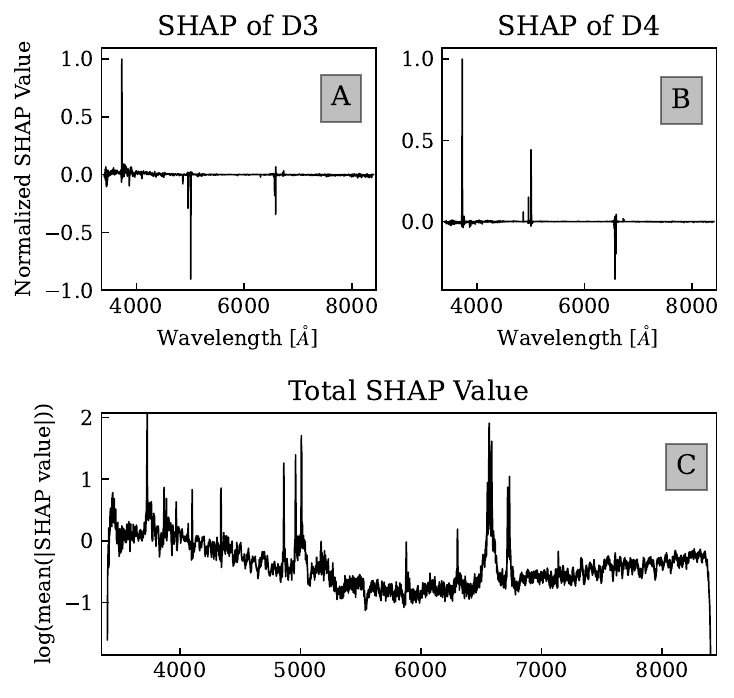}
        \caption{
        SHAP values as a function of wavelength for predicting latent variables. These figures display SHAP values as a function of wavelength, representing the importance of each wavelength in predicting latent variables. Panel A displays the SHAP values for D$_3$ for the galaxy (Plate1173-Mjd52790-FiberID111). Panel B presents the SHAP values for D$_4$ for the same galaxy. Panel C shows the logarithmically averaged absolute SHAP values of four latent variable. 
        }
        \label{figure SHAP value}
    \end{minipage}\hfill
\end{figure*}

Figure \ref{figure BIC} shows that the CVAE model conditioned with $M_{*}$ and the model conditioned with SFR ranks higher in the metric, implying that $M_{*}$ and SFR contribute most significantly to describing the galaxy spectral distribution while penalizing for the increasing model complexity. Figure \ref{figure spectra change CVAE}  shows the spectra reconstructed by the CVAEs when changing each physical property while keeping other latent variables constant at zero. When altering SFR, the intensity of the reconstructed spectra below 5000\AA \ decreases while the continuum above 5000\AA \ remains mostly the same. As expected, we see significant variation of [O II], [O III], and H$\alpha$ emission lines with SFRs. This means that SFR is relatively independent of the continuum shape. However, that is not the case when $M_{*}$, sSFR, and metallicity are altered. $M_{*}$ and metallicity are expected to be correlated \citep[e.g.,][]{tremonti_origin_2004}. Therefore, increasing $M_{*}$ and metallicity show similar changes to the reconstructed spectra. Decreasing the sSFR shows the continuous change from spectra dominated by young stellar populations with prominent nebular emissions to older stellar population emissions with minimal nebular emissions. These changes are in line with spectra changes expected for star-forming to quiescent stages of a galaxy. 
These results show that SFR affects the reconstructed spectra differently from $M_{*}$, sSFR, and metallicity. 

\subsection{Which spectral ranges are the most informative for representing observed SEDs?}

Figure \ref{figure SHAP value} shows the SHAP values in predicting the latent variables as a function of wavelength. SHAP values provide a measure of the contribution of each feature in predicting the latent variables, thereby indicating their importance. Panel C displays the average importance of each input wavelength in predicting latent variables, as measured by average absolute SHAP values. By taking the absolute value of these SHAP values, we emphasize the magnitude of a feature's impact on the predictions
Our analysis reveals the follwing pattern: wavelengths above 5000\textup{\AA}, with the exception of a few specific emission lines, exhibit minimal influence on our predictions. Therefore, it appears that an adequate characterization of galaxy spectra relies primarily on the data below 5000\textup{\AA} and selected emission lines. 
\section{Conclusion}\label{conclusion}
This study employed VAE and CVAE to extract four fundamental parameters from high-dimensional galaxy spectra. Our aim was to assess how these parameters, representing physical properties, affect the reconstruction of galaxy spectra and the formation of informative latent features. Specifically, D$_{1}$ is associated with $M_{*}$, and D$_{1}$ correlates with the SFR. D$_{3}$ indicates the ratio of [O II] to [O III] and H$\alpha$ intensities, whereas D$_{4}$ represents the ratio of H$\alpha$ to [O III] and [O II] emissions. Our findings show that incorporating $M_{*}$ and SFR into the CVAE model enhances the accuracy of galaxy spectra reconstruction. Additionally, SHAP analysis identified that wavelengths below 5000\AA \ and certain emission lines are particularly influential in these spectral ranges.

\clearpage
\section*{Acknowledgments}\label{Acknowledgments}
We express our deep appreciation to the five anonymous reviewers whose thorough reviews and insightful suggestions have greatly improved the quality of this manuscript. SC is supported by the Japan Society for the Promotion of Science (JSPS) under Grant No. 21J23611. This work has been supported by JSPS Grants-in-Aid for Scientific Research (TT: JP19H05076). This work has been supported by the Japan Society for the Promotion of Science (JSPS) Grants-in-Aid for Scientific Research (21H01128 and JP21J23611). This work has also been supported in part by the Collaboration Funding of the Institute of Statistical Mathematics "New Perspective of the Cosmology Pioneered by the Fusion of Data Science and Physics". Additionally, We also acknowledge the Center for Computational Astrophysics at the National Astronomical Observatory of Japan for providing access to their GPU cluster, which significantly supported our study.
This research utilized several software packages, including Astropy \cite{the_astropy_collaboration_astropy_2018},  Matplotlib \cite{hunter_matplotlib_2007}, NumPy \cite{oliphant_numpy_2006}, PyTorch \cite{paszke_pytorch_2019}, scikit-learn \cite{pedregosa_scikit-learn_2011}.
\section*{Checklist for Authors}
\begin{enumerate}
    \item For all authors...
    \begin{enumerate}[label=(\alph*)]
        \item Do the main claims made in the abstract and introduction accurately reflect the paper’s contributions and scope? \textcolor{blue}{[Yes]}
        \item Did you describe the limitations of your work? \textcolor{blue}{[Yes]}
        
        See Chapter \ref{discussion} for a discussion on the limited architectural experimentation and the analysis conducted on a single dataset. Further details will be elaborated in the main paper, which is being prepared for submission to an astronomical journal.
        \item Did you discuss any potential negative societal impacts of your work? \textcolor{gray}{[N/A]}
        \item Have you read the ethics review guidelines and ensured that your paper conforms to them? \textcolor{blue}{[Yes]}
    \end{enumerate}

    \item If you are including theoretical results...
    \begin{enumerate}[label=(\alph*)]
        \item Did you state the full set of assumptions of all theoretical results \textcolor{gray}{[N/A]}
        \item Did you include complete proofs of all theoretical results? \textcolor{gray}{[N/A]}
    \end{enumerate}

    \item If you ran experiments...
    \begin{enumerate}[label=(\alph*)]
        \item Did you include the code, data, and instructions needed to reproduce the main experimental results (either in the supplemental material or as a URL)? 

        \textcolor{blue}{[Yes]} All the code for downloading spectra, preprocessing, the model, training the model, and creating the figures used in this paper is publicly available in my GitHub repository: \faGithub\href{https://github.com/iwasakida}{iwasakida}. 
        \item Did you specify all the training details (e.g., data splits, hyperparameters, how they were chosen)?

        \textcolor{blue}{[Yes]}, Training details are also publicly available
        \item Did you report error bars (e.g., with respect to the random seed after running experiments multiple times)? \textcolor{orange}{[No]}
        
        \item Did you include the total amount of compute and the type of resources used (e.g., type of GPUs, internal cluster, or cloud provider)?
        
        \textcolor{blue}{[Yes]}, refer to Section \ref{Acknowledgments} for details.
    \end{enumerate}

    \item If you are using existing assets (e.g., code, data, models) or curating/releasing new assets...
    \begin{enumerate}[label=(\alph*)]
        \item If your work uses existing assets, did you cite the creators? \textcolor{blue}{[Yes]}
        \item Did you mention the license of the assets? 

        \textcolor{orange}{[No]} I used publicly accessible Galaxy data from SDSS with no specific license
        \item Did you include any new assets either in the supplemental material or as a URL? \textcolor{gray}{[N/A]}
        \item Did you discuss whether and how consent was obtained from people whose data you’re
using/curating?

        \textcolor{blue}{[Yes]} The process of downloading SDSS data details is also publicly available.
        \item Did you discuss whether the data you are using/curating contains personally identifiable information or offensive content? \textcolor{gray}{[N/A]}
    \end{enumerate}

    \item If you used crowdsourcing or conducted research with human subjects...
    \begin{enumerate}[label=(\alph*)]
        \item Did you include the full text of instructions given to participants and screenshots, if applicable? \textcolor{gray}{[N/A]}
        \item Did you describe any potential participant risks, with links to Institutional Review Board (IRB) approvals, if applicable? \textcolor{gray}{[N/A]}
        \item Did you include the estimated hourly wage paid to participants and the total amount spent on participant compensation? \textcolor{gray}{[N/A]}
    \end{enumerate}
    
\end{enumerate}
\medskip
{
\small
\bibliographystyle{apalike}
\bibliography{references}
}
\appendix{
\section{VAE} \label{VAE}
The main objective of VAE is to approximate the posterior distribution of latent variables given the input data. In VAE, first, we determine the architecture of the neural network (Figure \ref{figure architecture}) and then find the best parameters to approximate the probability distribution. This is done by maximizing the Evidence Lower Bound (ELBO). The ELBO is expressed as:
\begin{equation}
   \mathcal{L}(\theta, \phi; \bm{x}) = \mathbb{E}_{q_{\phi}(\bm{z}|\bm{x})}[\log p_{\theta}(\bm{x}|\bm{z})] - D_{KL}(q_{\phi}(\bm{z}|\bm{x}) \| p(\bm{z}))
\end{equation}
   where \(\mathbb{E}\) denotes the expectation, \(D_{KL}\) is the Kullback-Leibler divergence, \(\theta\) and \(\phi\) are neural network parameters, \(\bm{z}\) are latent variables, and \(\bm{x}\) is the observed data. The first term \(\mathbb{E}_{q_{\phi}(\bm{z}|\bm{x})}[\log p_{\theta}(\bm{x}|\bm{z})]\) is the reconstruction loss, which encourages the decoded samples to be close to the original inputs. The second term \(D_{KL}(q_{\phi}(\bm{z}|\bm{x}) \| p(\bm{z}))\) is the KL divergence, which measures the difference between the learned distribution \(q_{\phi}(\bm{z}|\bm{x})\) and the prior distribution \(p(\bm{z})\), typically a standard normal distribution. Each dimension is independent $p(\bm{z})=\Pi_j p(z_j)$. Thus, this regularization constrains each element of the latent variable to be independent, which allows us to get disentangle representation. A VAE not only learns the reconstruction but also the representation $z\sim q_{\phi}(\bm{z}|\bm{x})$. In deep generative models, representation learning is equivalent to inference. That is why VAE is known as a good method for representation learning and we use it. 
   
   To enable gradient-based optimization, VAE uses the reparameterization trick. Latent variables are expressed as a deterministic function of the input and some random noise:
\begin{equation}
   \bm{z} = \bm{\mu} + \bm{\sigma} \odot \bm{\epsilon}, \quad \bm{\epsilon} \sim \mathcal{N}(0, \bm{I})
\end{equation}
   where \(\bm{\mu}\) and \(\bm{\sigma}\) are the mean and standard deviation of the latent distribution predicted by the encoder, and \(\odot\) represents element-wise multiplication.

The VAE loss function combines the reconstruction loss (typically Mean Squared Error for continuous data) and the KL divergence:
\begin{equation}
   \mathcal{L}(\theta, \phi; \bm{x}) = \frac{1}{L} \sum_{l=1}^{L} \log p_{\theta}(\bm{x}|\bm{z}^{(l)}) - D_{KL}[q_{\phi}(\bm{z}|\bm{x}) \| p(\bm{z})]
\end{equation}
   with \(\bm{z}^{(l)}\) being samples drawn from the latent distribution using the reparameterization trick.
During training, a VAE aims to minimize the negative ELBO by adjusting its parameters, \(\theta\) and \(\phi\). This approach is effectively equivalent to maximizing the ELBO. Through this process, the VAE trains the encoder to generate meaningful latent representations and the decoder to accurately reconstruct the input data from these representations.

\begin{figure*}
    \centering
    \includegraphics[width=\textwidth]{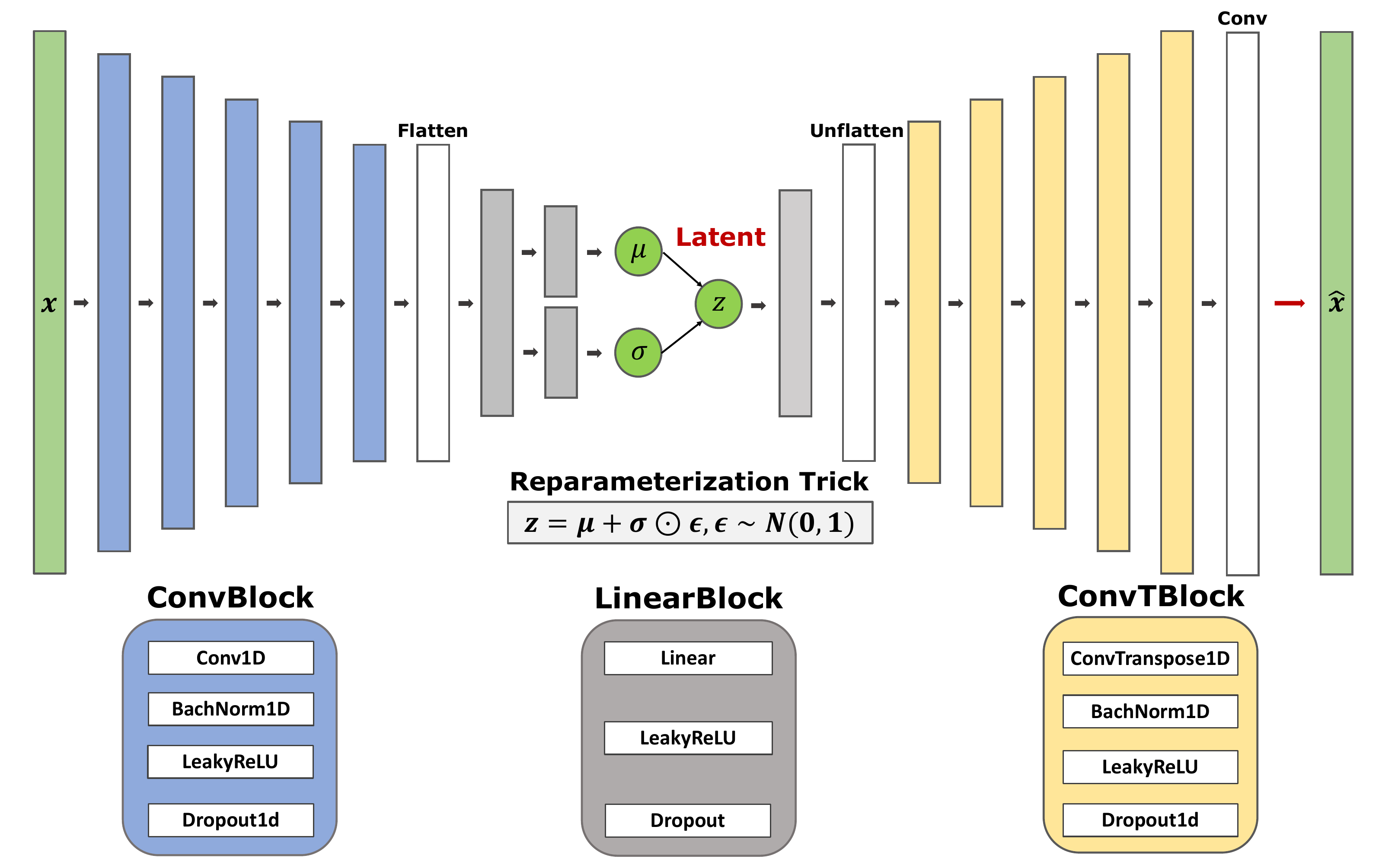}
    \caption{
        The illustration of the VAE architecture. The VAE architecture is depicted using color-coded blocks for different operations: ConvBloc in blue (Conv1D, BatchNorm1D, Dropout1D), LinearBlock in gray (Linear, LeakyReLU, Dropout), and ConvTBlock in yellow (ConvTranspose1D, BatchNorm1D, LeakyReLU, Dropout1D). Central white blocks represent Flatten and Unflatten operations for reshaping data, and an additional white block combines Conv1D, nn.ReLU, Flatten, Linear, and ReLU for matching input shape. These components collectively facilitate encoding of input \(\bm{x}\), decoding of output \(\bm{\hat{x}}\), and generation of patterns \(\bm{z}\). Key processes include feature extraction, data processing, reconstruction, and data reshaping, with the Reparameterization Trick introducing learnable parameters \(\bm{\mu}\) and \(\bm{\sigma}\) for the latent space distribution \(\bm{z}\).
    }
    \label{figure architecture}
\end{figure*}
\section{The SHAP values analysis} \label{The SHAP values analysis}
We employ the DeepExplainer function from Python's \texttt{shap} library to analyze SHAP values. These values, based on cooperative game theory, explain the output of machine learning models like our encoder $g_\phi$. They measure the significance of each feature in a prediction by comparing its impact to a baseline value. This analysis reveals how individual features influence the model's decision-making. In a model $f$ (our Encoder $g_\phi$), with $M$ input features, the SHAP value $\phi_i$ for feature $i$ quantifies this influence based on cooperative game theory principles.The SHAP value $\phi_i(x)$ is calculated using the formula:
\begin{equation}
\phi_i(x) = \sum_{S \subseteq N \setminus {i}} \frac{{|S|!(M - |S| - 1)!}}{M!} [f_x(S \cup {i}) - f_x(S)],
\end{equation}
where $f_x(S)$ is the model output with a specific set of features $S$, and $N$ is the total feature set. The formula considers all possible subsets of features excluding feature $i$, calculating the change in model output when $i$ is included versus excluded. The term $\frac{{|S|!(M - |S| - 1)!}}{M!}$ weights these changes to reflect the numerous combinations of features. This method quantifies the importance of each feature in the model's prediction, ensuring a fair distribution of impact among features, a concept rooted in cooperative game theory.
}
\end{document}